\documentclass[aps,prl,preprint,superscriptaddress]{revtex4-2}
\usepackage{graphicx}
\usepackage{bm}

\def\beq{\begin{equation}}
\def\eeq{\end{equation}}

\begin{document}
\title{Jahn-Teller magnets}
\author{Alexander Moskvin}
\affiliation{Ural Federal University, 620083 Ekaterinburg, Russia}


\begin{abstract}
A wide class of materials  with different crystal and electronic structures from quasi-two-dimensional unconventional superconductors (cuprates, nickelates, ferropnictides/chalcogenides, ruthenate SrRuO$_4$), 3D systems as manganites RMnO$_3$, ferrates (CaSr)FeO$_3$, nickelates RNiO$_3$, to silver oxide AgO  are based on Jahn-Teller $3d$ and $4d$ ions. These unusual materials called Jahn-Teller (JT) magnets are characterized by an extremely rich variety of phase states from non-magnetic and magnetic insulators to unusual metallic and superconducting states.
The unconventional properties of the JT-magnets can be related to the instability of their highly symmetric Jahn-Teller "progenitors" with the ground orbital $E$-state to charge transfer with anti-Jahn-Teller $d$-$d$ disproportionation and the formation of a system of effective local composite spin-singlet or spin-triplet, electronic or  hole $S$-type bosons moving in a non-magnetic or magnetic lattice.
We consider specific features of the anti-JT-disproportionation reaction, properties of the electron-hole dimers, possible phase states of JT-magnets, effective Hamiltonians for single- and two-band JT-magnets, and present a short overview of physical properties for actual JT-magnets.
\end{abstract}

\maketitle

\section{Introduction}


We call Jahn-Teller (JT) magnets compounds based on  Jahn-Teller 3$d$- and 4$d$-ions with configurations of the $t_{2g}^{n_1}e_g^{n_2}$ type in a highly symmetrical octahedral, cubic, or tetrahedral environment and with ground state orbital $E$-doublet.
  These are compounds based on tetra-complexes with the  configuration $d^1$ (Ti$^{3+}$, V$^{4+}$), low-spin (LS)  configuration $d^3$ (V$^{2+}$, Cr$^{3+}$, Mn$^{4+}$), and high-spin (HS) configuration $d^6$ (Fe$^{2 +}$, Co$^{3+}$), octa-complexes with HS-configuration $d^4$ (Cr$^{2+}$, Mn$^{3+}$, Fe$^{4+}$, Ru$^{4 +}$), low-spin configuration $d^7$ (Co$^{2+}$, Ni$^{3+}$, Pd$^{3+}$), as well as octa-complexes with configuration $d^9$ (Cu$^{2+}$, Ni$^{1+}$, Ag$^{2+}$) (see Table\,I). The class of JT-magnets includes a large number of promising materials that are the focus of modern condensed matter physics, such as RMnO$_3$ manganites, (Ca,Sr)FeO$_3$ ferrates, ruthenates RuO$_2$, (Ca,Sr)RuO$_3$, (Ca,Sr)$_2$RuO$_4$, a wide range of ferropnictides (FePn) and ferrochalcogenides (FeCh), RNiO$_3$ 3D nickelates, 3D- cuprate KCuF$_3$, 2D cuprates (La$_2$CuO$_4$, ...) and nickelates RNiO$_2$, silver-based compounds (AgO, AgF$_2$) (see Table\,I ).
  
Among these materials, it is necessary to highlight JT magnets with charge transfer, in particular, disproportionation, which have a rich  spectrum of unique properties from various types of magnetic and charge ordering to metal-insulator transitions and superconductivity.
Interestingly, the choice of cuprates as promising superconducting materials and the discovery of high-temperature superconductivity (HTSC)\,\cite{HTSC} were stimulated precisely by the outstanding Jahn-Teller character of Cu$^{2+}$ ions.

Many authors considered disproportionation as a mechanism ("negative-$U$"\,model) leading to "glueless" superconductivity of a system of local electron pairs, or composite bosons (see, e.g., Refs.\,\cite{Ionov_1975,Anderson_1975,Scheurell,Larsson_2003,Wilson,Hirsch_1985,Sleight,Kulik,Rice_1981,David,Varma_1988,Dzyalo_2,Geballe,Mitsen,Tsendin,Katayama,Pouchard,Mazin}).
The concept of superconductivity as a Bose-Einstein condensation (BEC) of local composite bosons as two electrons bound in real space was introduced by Ogg Jr. in 1946\,\cite{Ogg} and developed by Schafroth in 1954-55\,\cite{Schafroth}. However, due to the triumph of the BCS (Bardeen-Cooper-Schrieffer) theory, the idea of local composite bosons and preformed pairs was practically forgotten for many years. The discovery of HTSC cuprates in 1986 revived interest in the idea of local pairing\,\cite{Alexandrov_2011}, especially since this idea has been supported by K.~A. Mueller, the discoverer of HTSC\,\cite{Mueller}. Now there is convincing experimental evidence that the local pairing of carriers  takes place well above T$_c$ at least in underdoped cuprates\,\cite{Pavuna}.
At the same time, until now, one way or another, the HTSC theory has been dominated by the approaches based on the BCS paradigm, i.e, on the representations of the BCS model theory applicable to the description of typical low-temperature superconductors.
This is largely due to the fact that an appealingly simple picture of the preformed pairs and BEC superconductivity in cuprates seemingly came to be at odds with a number of experimental observations of the typical Fermi liquid behavior, and notably with indications that
a well-defined Fermi surface (FS) exists, at least in overdoped cuprates, the thermal and electrical conductivity were found to follow the standard Wiedemann-Franz law, quantum oscillations have been observed as well in various cuprates\,\cite{Bozovic}.

However, this contradictory behavior can be easily explained if we take into account the possibility of separating the superconducting BEC phase and the normal Fermi-liquid phase. Indeed, recently Pelc {\it et al}.\,\cite{Pelc} have introduced a phenomenological model of a "local phase separation" wherein two electronic subsystems coexist within the unit cell: itinerant and localized holes, with the $p$ holes introduced via doping always being itinerant while pairing is associated with the localized holes.
In fact, they argue that the Fermi liquid subsystem in cuprates is responsible for the normal state with angle-resolved photoemission spectra (ARPES), magnetic quantum oscillations, and Fermi arcs, but not for the unconventional superconducting state. In other words, {\it cuprate superconductivity is not related to the doped hole pairing}, the carriers which exhibit the Fermi liquid behaviour are not the ones that give rise to superconductivity. However, the authors could not elucidate the nature of local pairing to be a central point of the cuprate puzzle.


The disproportionation scenario, especially popular in the "chemical"\, community ("chemical"\, way to superconductivity), was addressed earlier by many authors, however, by now it was not properly developed theoretically, and perhaps that is why it has not yet been a worthy competitor to the traditional BCS approach.

Previously, we proposed a mechanism for anti-Jahn-Teller disproportionation\,\cite{dispro} with the formation of a system of local composite spin-triplet bosons and unconventional bosonic superconductivity in 3$d$ JT magnets.
  This mechanism indicated, in particular, the spin-triplet superconductivity of ferropnictides and ferrochalcogenides, which was predicted back in 2008\,\cite{FPS}. Over the past years, new results have been obtained in the study of JT magnets based on both $3d$ and $4d$ ions,  as well as new arguments both for and against spin-triplet superconductivity.

In this paper, we develop a model of "anti-Jahn-Teller"\ disproportionation into a wider class of Jahn-Teller magnets, including $4d$ magnets (ruthenates, silver compounds) and 2D nickelates RNiO$_2$, and show that all of them can be described within a single scenario. In Sec.\,II, we present a more detailed description of the anti-JT disproportionation for JT-magnets and formation of effective local composite bosons. In Secs.\,III and IV we consider electron-hole (EH) dimers as specific "disproportionation quanta", their electron and spin structure. Sec.\,V provides a brief overview of the possible phase states for JT magnets. Secs. VI and VII present the effective Hamiltonians of single- and two-band JT magnets with a brief overview of the properties of real JT magnets.
 A brief conclusion is given in Sec.\,VIII.


\section{Anti-Jahn-Teller disproportionation }

The lifting of the orbital $E$-degeneracy in the high-symmetry "progenitor"\, JT-magnets can be associated both with the specifics of the crystal structure, as, for example, in "apex-free"\, 2D cuprates (Nd$_2$CuO$_4$) and RNiO$_2$ nickelates, and with the conventional Jahn-Teller effect, which, as a rule, leads to the formation of a low-symmetry  insulating  antiferromagnetic (La$_2$CuO$_4$, KCuF$_3$, LaMnO$_3$) or ferromagnetic (K$_2$CuF$_4$) phase. A competing mechanism for removing orbital degeneracy in JT magnets considered above is "anti-Jahn-Teller", "symmetric"\, $d$-$d$-disproportionation according to the scheme
\beq
d^n + d^n \rightarrow d^{n+1} + d^{n-1} \, ,
\label{dis}
\eeq
assuming the formation of a system of bound or relatively free electronic $d^{n+1}$ and hole $d^{n-1}$ centers, differing by a pair of electrons/holes.
Formally, an electron/hole center can be represented as a hole/electron center with a pair of electrons/holes $d^2$/$\underline{d}^2$ localized at the center. In other words, a disproportionate system can be formally represented as a system of effective local spin-singlet or spin-triplet composite electron/hole bosons "moving"\, in the lattice of hole/electron centers. Note that in frames of the toy model (\ref{dis}) the disproportionation energy $\Delta_{dd}$ formally coincides with the energy of local correlations $U_{dd}$, which gives reason to associate symmetric $d$-$d$-disproportionation with the negative-$U$ phenomenon.

Obviously, in systems with strong $d$-$p$-hybridization (cation-anion covalence), the disproportionation reaction (\ref{dis})
must be written in a "cluster" language, for example, as for CuO$_4$ clusters in the CuO$_2$ cuprate planes
\beq
[\mbox{CuO}_4]^{6-}+[\mbox{CuO}_4]^{6-}\,\rightarrow\,[\mbox{CuO}_4]^{7-}+[\mbox {CuO}_4]^{5-}\, ,
\label{disCuO}
\eeq
instead of
\beq
d^9 + d^9 \rightarrow d^{10} + d^{8}  \, .
\eeq
The cation-ligand cluster representation of the $d^n,d^{n\pm 1}$-centers immediately indicates the important role of the so-called "breathing mode" of the ligand displacements in perovskite-type JT-magnets with corner-shared coupling of neighboring octahedral $d$-centers.
The displacement amplitude of the common ligand for two centers during disproportionation can reach values greater than  of 0.1\,\AA\, due to the large difference in the cation–ligand separation for the electron and hole centers. Thus, the Cu-O separation for CuO$_4$-centers in cuprates increases by 0.2\,\AA\, from the hole [CuO$_4$]$^{5-}$ to electron [CuO$_4$]$^{7-}$ center\,\cite{Larsson_2007}. Softening of the breathing mode is considered an indication of disproportionation.
The electron-lattice interaction  leads to the stability of the electron and hole centers in the lattice of the parent system with the ground states of all three centers -- the electron, parent, and hole one   corresponding to different values of the local breathing configuration coordinate $Q_{A_{1g}}$: $+Q_0$, 0, $-Q_0$, respectively.

Note that in any case, "symmetric"\, $d$-$d$-disproportionation, in contrast to "asymmetric",\, "single-center",\, $d$-$p$-disproportionation\,\cite{Vikhnin,Mazumdar,Sawatzky} has a two-center character, although it may include $d$-$p$-transfer between clusters. Obviously, symmetric $d$-$d$ disproportionation will be energetically more favorable in "progenitor"\, Mott-Hubbard JT magnets, and vice versa, asymmetric $d$-$p$-disproportionation will be more energetically favorable in charge-transfer (CT) insulators ("negative charge transfer"\, materials).

It is worth noting that all the JT-magnets are characterized by empty, half-filled or fully filled $t_{2g}$-subshell with orbitally non-degenerate, or $S$-type ground state, and with only one $e_g$-electron or hole. Obviously, the anti-JT-disproportionation implies the $e_g$-$e_g$ intersite transfer with the formation of the empty, half-filled or fully filled $e_{g}$-subshells with the $S$-type ground state for the electron and hole centers. In all cases, we arrive at relatively stable $S$-type configurations of electron and hole centers. For all JT magnets, the anti-JT disproportionation reactions can be written as follows
\begin{eqnarray}
& tetra\,\, d^1:   e_g^1+ e_g^1 \stackrel{e_g}{\rightarrow} \left\{\begin{array}{c}
{\bf e_g^0} + {\bf e_g^0}e_g^2 \\
{\bf e_g^0}e_g^2+{\bf e_g^0} \\
\end{array}\right\} \, ; \\
& tetra\,\, d^3:   e_g^3+ e_g^3 \stackrel{e_g}{\rightarrow} \left\{\begin{array}{c}
{\bf e_g^4} + {\bf e_g^4}\underline{e}_g^2 \\
{\bf e_g^4}\underline{e}_g^2+{\bf e_g^4} \\
\end{array}\right\} \, ; \\
& octa \,\, d^4:t_{2g}^3e_g +t_{2g}^3e_g \stackrel{e_g}{\rightarrow} \left\{\begin{array}{c}
{\bf t_{2g}^3} + {\bf t_{2g}^3}e_g^2  \\
{\bf t_{2g}^3}e_g^2 + {\bf t_{2g}^3}  \\
\end{array}\right\} \, ; \\
& tetra \,\, d^6:e_g^3t_{2g}^3 +e_g^3t_{2g}^3 \stackrel{\underline{e}_g}{\rightarrow} \left\{\begin{array}{c}
{\bf e_g^4t_{2g}^3}\underline{e}_g^2 +{\bf e_g^4t_{2g}^3} \\
{\bf e_g^4t_{2g}^3}+{\bf e_g^4t_{2g}^3}\underline{e}_g^2 \\
\end{array}\right\} \, ;  \\
& octa \,\, d^7:t_{2g}^6e_g +t_{2g}^6e_g \stackrel{e_g}{\rightarrow} \left\{\begin{array}{c}
{\bf t_{2g}^6} + {\bf t_{2g}^6}e_g^2  \\
{\bf t_{2g}^6}e_g^2 +{\bf t_{2g}^6}  \\
\end{array}\right\} \, ; \\
& octa \,\, d^9:t_{2g}^6e_g^3 +t_{2g}^6e_g^3 \stackrel{\underline{e}_g}{\rightarrow} \left\{\begin{array}{c}
{\bf t_{2g}^6e_g^4}\underline{e}_g^2 + {\bf t_{2g}^6e_g^4}  \\
{\bf t_{2g}^6e_g^4}+{\bf t_{2g}^6e_g^4}\underline{e}_g^2   \\
\end{array}\right\} \, \, .
\label{dis1}
\end{eqnarray}

Obviously, for JT magnets with the on-site progenitor configurations $e_g^1$, $t_{2g}^3e_g^1$, $t_{2g}^6e_g^1$ we are dealing with the transfer of the $e_g$-electron, while for configurations $e_g^3t_{2g}^3$ and $t_{2g}^6e_g^3$ it is correct to speak of the $e_g$-hole ($\underline{e}_g$) transfer. Thus for these  configurations we arrive at a doublet of  ionic states with site-centred charge order,
or two centers that differ in the transfer (exchange) of two $e_g$-electrons or two $e_g$-holes, respectively,
 that can be thought of as  effective  local composite bosons. For the centers with high (octahedral, tetrahedral) symmetry these effective bosons  will be described by the low-energy Hund configuration $e_g^2;{}^3A_{2g}$ or $\underline{e}_g^2;{}^3A_{2g}$.
 It should be noted that effective bosons cannot be considered as conventional quasiparticles, they are an indivisible part of many-electron configurations.

\begingroup
\begin{table*}
\centering
\caption{$3d^n$ and $4d^n$ JT-magnets.  }
\begin{tabular}{|c|c|c|c|c|c|}
\hline
  \begin{tabular}{c}
JT-configuration \\
JT ions \\
\end{tabular}       & Symm. & LS/HS & \begin{tabular}{c}
Local \\
boson \\
\end{tabular}  & Lattice& \begin{tabular}{c}
Representative \\
compounds \\
\end{tabular}   \\ \hline
  \begin{tabular}{c}
$3d^1$($e_g^1$):${}^2E$ \\
Ti$^{3+}$, V$^{4+}$, Cr$^{5+}$ \\
\end{tabular}  & tetra& -&\begin{tabular}{c}
$e_g^2$:${}^3A_{2g}$ \\
s=1 \\
\end{tabular}  & \begin{tabular}{c}
A$_{1g}$ \\
S=0 \\
\end{tabular} &\begin{tabular}{c}
$\beta$-Sr$_2$VO$_4$ \\
(Sr,Ba)$_3$Cr$_2$O$_8$  \\
\end{tabular} \\ \hline
  \begin{tabular}{c}
$3d^3$($e_g^3$):${}^2E$ \\
V$^{2+}$, Cr$^{3+}$, Mn$^{4+}$ \\
\end{tabular}  & tetra& LS &\begin{tabular}{c}
$\underline{e}_g^2$:${}^3A_{2g}$ \\
s=1 \\
\end{tabular}  & \begin{tabular}{c}
A$_{1g}$ \\
S=0 \\
\end{tabular} & Ba$_2$VGe$_2$O$_7$ (?) \\ \hline
  \begin{tabular}{c}
$3d^4$($t_{2g}^3e_g^1$):${}^5E$ \\
Cr$^{2+}$, Mn$^{3+}$, Fe$^{4+}$ \\
\end{tabular}& octa & HS &\begin{tabular}{c}
$e_g^2$:${}^3A_{2g}$ \\
s=1 \\
\end{tabular}   & \begin{tabular}{c}
A$_{2g}$ \\
S=3/2 \\
\end{tabular} & \begin{tabular}{c}
CrO, CrF$_2$ \\
Sr$_2$FeO$_4$ \\
(Ca,Sr,Ba)FeO$_3$\\
(Ca,Sr,Ba)$_3$Fe$_2$O$_7$ \\
RMnO$_3$, LaMn$_7$O$_{12}$ \\
\end{tabular} \\  \hline
 \begin{tabular}{c}
$4d^4$($t_{2g}^3e_g^1$):${}^5E$ \\
Ru$^{4+}$ \\
\end{tabular}& octa & HS &\begin{tabular}{c}
$e_g^2$:${}^3A_{2g}$ \\
s=1 \\
\end{tabular}   & \begin{tabular}{c}
A$_{2g}$ \\
S=3/2 \\
\end{tabular} & \begin{tabular}{c}
(Ca,Sr)$_2$RuO$_4$\\
(Ca,Sr)RuO$_3$, RuO$_2$\\
(Ca,Sr)$_3$Ru$_2$O$_7$\\
\end{tabular} \\   \hline
  \begin{tabular}{c}
$3d^6$($e_g^3t_{2g}^3$):${}^5E$ \\
Fe$^{2+}$, Co$^{3+}$ \\
\end{tabular}& tetra& HS&\begin{tabular}{c}
$\underline{e}_g^2$:${}^3A_{2g}$ \\
s=1 \\
\end{tabular}   & \begin{tabular}{c}
A$_{2g}$ \\
S=3/2 \\
\end{tabular} & FePn, FeCh, Na$_5$CoO$_4$     \\ \hline
  \begin{tabular}{c}
$3d^7$($t_{2g}^6e_g^1$):${}^2E$ \\
Co$^{2+}$, Ni$^{3+}$ \\
\end{tabular}& octa& LS&\begin{tabular}{c}
$e_g^2$:${}^3A_{2g}$ \\
s=1 \\
\end{tabular}  & \begin{tabular}{c}
A$_{1g}$ \\
S=0 \\
\end{tabular} & \begin{tabular}{c}
RNiO$_3$\\
(Li,Na,Ag)NiO$_2$ \\
\end{tabular} \\ \hline
 \begin{tabular}{c}
$3d^9$($t_{2g}^6e_g^3$):${}^2E$ \\
Cu$^{2+}$, Ni$^+$ \\
\end{tabular}&
octa  & -&\begin{tabular}{c}
$\underline{e}_g^2$:${}^3A_{2g}$ \\
s=1 \\
\end{tabular}  & \begin{tabular}{c}
A$_{1g}$ \\
S=0 \\
\end{tabular} &
CuF$_2$, KCuF$_3$, K$_2$CuF$_4$\\
\hline
\begin{tabular}{c}
$4d^9$($t_{2g}^6e_g^3$):${}^2E$ \\
Pd$^+$, Ag$^{2+}$ \\
\end{tabular}&octa& -&\begin{tabular}{c}
$\underline{e}_g^2$:${}^3A_{2g}$ \\
s=1 \\
\end{tabular}  & \begin{tabular}{c}
A$_{1g}$ \\
S=0 \\
\end{tabular} & AgO (Ag$^{1+}$Ag$^{3+}$O$_2$)    \\  \hline
  \begin{tabular}{c}
$3d^9$($t_{2g}^6e_g^3$):${}^2B_{1g}$ \\
Cu$^{2+}$, Ni$^+$ \\
\end{tabular}&\begin{tabular}{c}
octa$^*$\\
square \\
\end{tabular}  & -&\begin{tabular}{c}
$\underline{b}_{1g}^2$:${}^1A_{1g}$ \\
s=0 \\
\end{tabular}  & \begin{tabular}{c}
A$_{1g}$ \\
S=0 \\
\end{tabular} & \begin{tabular}{c}
HTSC cuprates \\
RNiO$_2$, CuO\\
\end{tabular}    \\ \hline
 \begin{tabular}{c}
$4d^9$($t_{2g}^6e_g^3$):${}^2B_{1g}$ \\
Pd$^+$, Ag$^{2+}$ \\
\end{tabular}& \begin{tabular}{c}
octa$^*$\\
square \\
\end{tabular}  & -&\begin{tabular}{c}
$\underline{b}_{1g}^2$:${}^1A_{1g}$ \\
s=0 \\
\end{tabular}  & \begin{tabular}{c}
A$_{1g}$ \\
S=0 \\
\end{tabular} & \begin{tabular}{c}
AgF$_2$, KAgF$_3$ \\
Cs$_2$AgF$_4$, LaPdO$_2$ (?)  \\
\end{tabular}     \\ \hline
       \end{tabular}
\end{table*}
\endgroup

 All JT-magnets can be conditionally divided into the "single-band" and "two-band" ones. In single-band JT-magnets with the configurations $d^1$, $d^3$, $d^7$, and $d^9$, effective electron ($d^1$, $d^7$) or hole ($d^3$, $d^9$) composite bosons move in the lattice of ions with completely filled shells, while in two-band JT-magnets ($d^4$, $d^6$) the lattice includes ions with half filled  $t_{2g}$-subshell.

The optimal configurations and spin of the composite boson, as well as the orbital state and local spin of the lattice, formed as a result of anti-JT disproportionation in JT magnets with the 3$d^n$ configuration, as well as some 4$d^n$ JT configurations, are given in the fourth and fifth columns of Table\,I. Note that in all cases the complete disproportionation leads to a system of composite bosons with the concentration of 1/2, or half-filling.

\section{Electron-hole dimers}

Pair of bound electron and hole centers, an EH dimer, is a kind of "disproportionation quantum". In Mott-Hubbard insulators, EH dimers are low-energy metastable charge excitations above the ground state or may be the result of the self-trapping of the $d$-$d$ CT-excitons\,\cite{OS_2023}.

 The two electron/hole charge exchange reaction in EH dimer
 \begin{equation}
d_1^{n+1}+d_2^{n-1}   \stackrel{e_g^2;{}^3A_{2g}}{\leftrightarrow}       d_1^{n-1}+d_2^{n+1} \, ,	
\label{12-21}
\end{equation}
is controlled by the effective local boson transfer integral
\begin{equation}
t_B=\langle d_1^{n+1}d_2^{n-1}|\hat H_B|d_1^{n-1}d_2^{n+1}\rangle \, ,	
\end{equation}
 where $\hat H_B$ is an effective two-particle (bosonic) transfer Hamiltonian and we assume  ferromagnetically ordered spins of the two centers.
As a result of this quantum process the bare ionic states with site-centered charge order and the same bare energy $E_0$ transform into two EH-dimer states with an indefinite valence and bond-centred charge-order
\begin{equation}
|\pm \rangle =\frac{1}{\sqrt{2}}(|d_1^{n+1}d_2^{n-1}\rangle \pm |d_1^{n-1}d_2^{n+1}\rangle )	
\end{equation}
with the energies $E_{\pm}=E_0\pm t_B$. In other words, the exchange reaction restores the bare charge symmetry.
 In both $|\pm\rangle $ states the on-site number of $d$-electrons is indefinite with quantum  fluctuations between ($n+1$) and ($n-1$), however, with a mean value $n$.
Interestingly that, in contrast with the  ionic states, the EH-dimer states $|\pm \rangle $  have both a distinct electron-hole and inversion symmetry, even parity ($s$-type symmetry) for $|+ \rangle $, and odd parity ($p$-type symmetry) for $|- \rangle $ states, respectively. The both states are coupled by a large electric-dipole matrix element:
\begin{equation}
\langle +|\hat {\bf d}|-\rangle =2eR_{12}\, , 	
\end{equation}
where $R_{12}$ is a $1-2$ separation.
The two-particle transport
(\ref{12-21}) can be realized through two successive one-particle processes with the $e_g$-electron transfer  as follows
$$
d_1^{n+1}+d_2^{n-1} \stackrel{e_g}{\rightarrow}  d_1^{n}+d_2^{n} \stackrel{e_g}{\rightarrow} d_1^{n-1}+d_2^{n+1}\, ,
$$
hence the two-particle transfer integral $t_B$ can be evaluated as follows:
\begin{equation}
t_B=-t_{e_ge_g}^2/U \approx -J_{kin}(e_ge_g)\,,	
\end{equation}
where $t_{e_ge_g}$ is one-particle transfer integral for $e_g$ electron, $U$ is a mean transfer energy. It means that  the two-particle bosonic transfer integral  can be directly coupled with the kinetic $e_g$-contribution $J_{kin}(e_ge_g)$ to Heisenberg $e_g$-$e_g$ exchange integral. Both $t_B$ and $J_{kin}(e_ge_g)$ are determined by the second order one-particle transfer mechanism.
It should be noted that negative sign of the two-particle CT integral $t_B$ points to the energy stabilization of the $s$-type EH-dimer state $|+ \rangle $.

Second, one should emphasize once more that the stabilization of EH-dimers is provided by a strong electron-lattice effect with a striking intermediate oxygen atom polarization and  displacement concommitant with charge exchange. In a sense, the EH-dimer may be addressed to be a bosonic counterpart of the Zener Mn$^{4+}$-Mn$^{3+}$ polaron\,\cite{Zener}. It is no wonder that even in a generic disproportionated system BaBiO$_3$ instead of simple checkerboard charge ordering of Bi$^{3+}$ and Bi$^{5+}$ ions we arrive at a CDW (charge density wave) state with the alteration of expanded Bi$^{(4-\rho )+}$O$_6$ and compressed Bi$^{(4+\rho )+}$O$_6$ octahedra with $0<\rho \ll 1$\,\cite{BaBiO-Merz}. Enormously large values of oxygen thermal parameters in BaBiO$_3$\,\cite{Chaillout} evidence a great importance of dynamical oxygen breathing modes providing some sort of a "disproportionation glue". Sharp rise of the oxygen thermal parameter in the high-temperature O phase of LaMnO$_3$\,\cite{Rodriguez} or in several "competing"\, phases found by Huang {\it et al.}\,\cite{Huang} as compared with the bare AFI phase is believed to be a clear signature of the high-temperature manganese disproportionation\,\cite{Moskvin-09}.

Actually we deal with an EH-dimer to be a dynamically charge fluctuating bipolaronic system of coupled electron $d^{n+1}$ and hole $d^{n-1}$ centers having been glued in a lattice due to a specific local expansion/contraction mode of neighboring clusters (half-breathing, or breathing mode) and strong electron-lattice polarization effects.



\section{Spin structure of EH-dimers}

Let's apply to spin degrees of freedom which are of great importance for magnetic properties of EH-dimers as nucleation centers for a rich variety of different phases.
First of all, we note that the structure of EH-dimers is significantly different in single- and two-band JT magnets. In EH-dimers of JT magnets based on $d^1$, $d^7$, and $d^9$ configurations, the spin-triplet boson "moves" along spinless centers (see Table 1), which leads to a trivial spin structure of the dimer. A more complicated situation is realized for EH-dimers of JT magnets based on  $d^4$ and  $d^6$ configurations, where the spin-triplet boson "moves" through the $d^3(t_{2g}^3)$-centers with spin 3/2 (see Table\,1).

The total spin moment of these EH-dimers is ${\bf S}={\bf S }_1+{\bf S}_2$, where ${\bf S }_1$ ($S_1=5/2$) and ${\bf S }_2$ ($S_1=3/2$) are spins of $d^5$ and $d^3$ ($d^5$ and $\underline{d}^3$)  configurations, respectively, so the total spin magnitude $S$  takes the values 1, 2, 3, 4.
In nonrelativistic approximation the spin structure of the EH-dimer in the bare ionic state $d^5$-$d^3$ ($d^3$-$d^5$) with site-centered charge order   will be determined by isotropic Heisenberg exchange coupling
\begin{equation}
V_{ex}=J(d^5d^3)\,({\bf S }_1\cdot {\bf S }_2),	
\end{equation}
 with $J(d^5d^3)$ being a $d^5$-$d^3$ ($d^3$-$d^5$) (super)exchange integral.
 However, the two site-centered states $d^5$-$d^3$ and $d^3$-$d^5$ are coupled by the two-particle charge transfer characterized by a respective transfer integral which depend on spin states as follows:
\begin{equation}
\langle \frac{5}{2}\frac{3}{2};SM|\hat H_B|\frac{3}{2}\frac{5}{2};SM\rangle =\frac{1}{20}S(S+1)\,t_B\, ,	
\label{tB}
\end{equation}
 where  $t_B$ is a spinless  transfer integral. Making use of this expression we can introduce an effective spin-operator form for the boson transfer as follows:
 \begin{equation}
	\hat H_B^{eff}=\frac{t_B}{20}\left[2(\hat {\bf S}_1\cdot\hat {\bf S}_2)+S_1(S_1+1)+S_2(S_2+1)\right]\, ,
 \label{tBS}
\end{equation}
which can be a very instructive tool both for qualitative and quantitative analysis of boson transfer effects.
Thus, the effective transfer integral of the composite boson strongly depends on the spin state of the electron-hole pair, falling tenfold as the total spin of the pair changes from $S$\,=\,4 to $S$\,=\,1.
In particular, we arrive at a strong, almost twofold, suppression of effective transfer integral in paramagnetic phase as compared with its maximal value $t_B$ for a ferromagnetic ordering ($S$\,=\,4).

  \begin{figure}[t]
 \begin{center}
\includegraphics[width=14cm,angle=0]{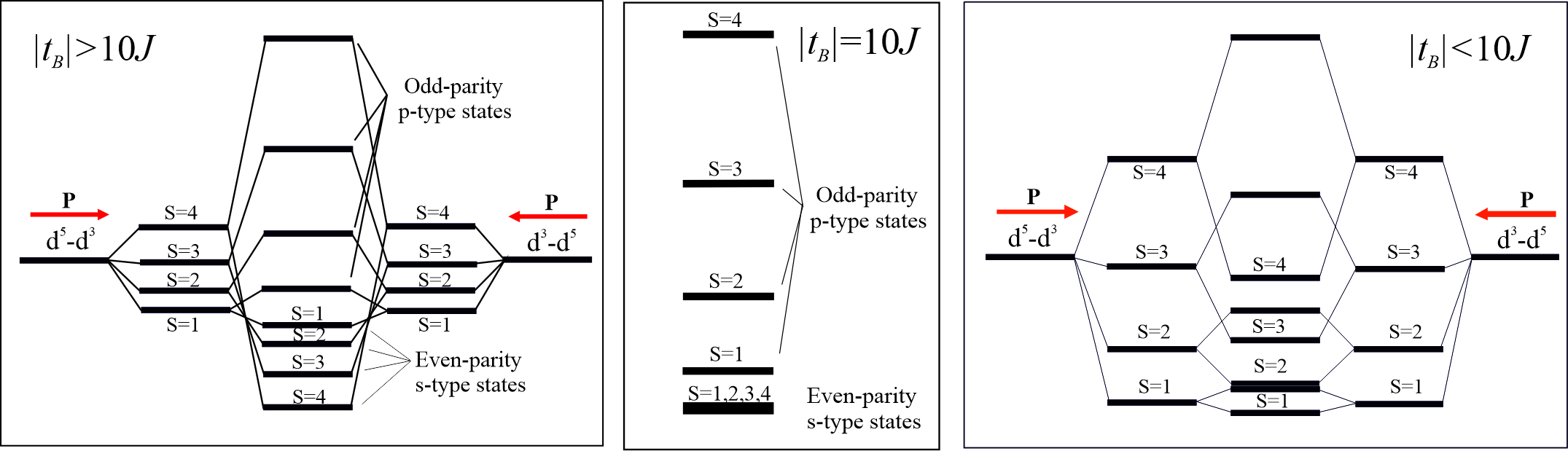}
\caption{(Color online) Spin structure of the EH-dimer, or self-trapped CT exciton with a step-by-step  inclusion of one- and two-particle charge transfer. Arrows point to electric dipole moment for bare site-centered dimer configurations.} \label{fig1}
\end{center}
\end{figure}

 Both conventional Heisenberg exchange coupling $d^5$-$d^3$ ($d^3$-$d^5$) and unconventional two-particle bosonic transfer, or bosonic double exchange can be easily diagonalized in the total spin $S$ representation so that for the energy of the EH-dimer we arrive at
\begin{equation}
E_S=\frac{J(d^5d^3)}{2}[S(S+1)-\frac{25}{2}]\pm \frac{1}{20}S(S+1)\,t_B\,,	
\end{equation}
where $\pm$ corresponds to two quantum superpositions $|\pm\rangle $ written in a spin representation as follows
\begin{equation}
|SM\rangle _{\pm} =\frac{1}{\sqrt{2}}(|\frac{5}{2}\frac{3}{2};SM\rangle \pm |\frac{3}{2}\frac{5}{2};SM\rangle )\, ,	
\end{equation}
with $s$- and $p$-type symmetry, respectively.  It is worth  noting that the bosonic double exchange contribution formally corresponds to ferromagnetic exchange coupling with $J_B=-\frac{1}{10}|t_B|$.

We see that the cumulative effect of the Heisenberg exchange and the bosonic double exchange results in a stabilization of the $S$\,=\,4 high-spin
 (ferromagnetic) state of the EH-dimer provided $|t_B|>10J$ (see left panel in Fig.\,\ref{fig1}) and  the $S$\,=\,1 low-spin ("ferrimagnetic") state otherwise (see right panel in Fig.\,\ref{fig1}). Spin states with intermediate S values: $S$\,=\,2, 3 correspond to a classical noncollinear ordering.
 It is interesting that for $|t_B|=10\,J$ the energy of the  dimer's  $S$-type states does not depend on the value of the total spin, so that we arrive at the surprising result of a 24-fold ($\sum_{S=1}^{S=4}(2S+1)$) degeneracy of the ground state of an isolated dimer (see central panel in Fig.\,1).

 To estimate the both quantities $t_B$ and $J(d^5d^3)$ and their dependence on the crystal structure parameters  we can address the results of a comprehensive theoretical and experimental analysis of different superexchange integrals in
  perovskites RFeO$_3$, RCrO$_3$, and RFe$_{1-x}$Cr$_x$O$_3$ with Fe$^{3+}$ and Cr$^{3+}$ ions with electronic configurations $d^5$ and $d^3$, respectively\,\cite{Ovanesyan,JETP_2021,MC_2021}. These perovskites are isostructural with many JT-magnets including (Ca,Sr,Ba)FeO$_3$, RMnO$_3$, (Ca,Sr,Ba)RuO$_3$.

Antiferromagnetic kinetic exchange contribution to $J(e_ge_g)$ related with the $e_g$-electron transfer to partially filled $e_g$-shell can be written as follows\,\cite{MC_2021,JETP_2021}
\begin{equation}
J(e_ge_g)=\frac{(t_{ss}+t_{\sigma\sigma}\cos\theta)^2}{2U} \, ,
\label{kinetic}	
\end{equation}
while for the $d^5$-$d^3$ superexchange we arrive at a competition of the antiferromagnetic and ferromagnetic contributions
$$
J_{FeCr}=J(d^5d^3)=\frac{2}{15}\left(\frac{t_{\sigma\pi}^2}{U}\sin^2\theta+\frac{t_{\pi\pi}^2}{U}(2-\sin^2\theta )\right)
$$
 \begin{eqnarray}
-\frac{\Delta E(35)}{10U}\left[\frac{(t_{ss}+t_{\sigma\sigma}\cos\theta)^2}{U}+\frac{t_{\sigma\pi}^2}{U}\sin^2\theta\right]    \,.
\end{eqnarray}
 Here $\theta$ is the cation-anion-cation bonding angle, $t_{\sigma\sigma}>t_{\pi\sigma}>t_{\pi\pi}>t_{ss}$ are positive definite $d$-$d$ transfer integrals, $U$ is a mean $d$-$d$ transfer energy (effective correlation energy), $\Delta E(35)$ is the energy separation between $^3E_g$ and $^5E_g$ terms for the $t_{2g}^3e_g$ configuration.

 Microscopically derived angular dependence of the superexchange integrals $J_{FeFe}$, $J_{CrCr}$, and $J_{FeCr}$ does nicely describe
 the full set of experimental data on the value of $T_N$ for various orthoferrites, orthochromites, mixed ferrites-chromites, as well as M\"{o}ssbauer data on Fe-substituted orthochromites\,\cite{Ovanesyan,MC_2021,JETP_2021}.

   \begin{figure}[t]
 \begin{center}
\includegraphics[width=8cm,angle=0]{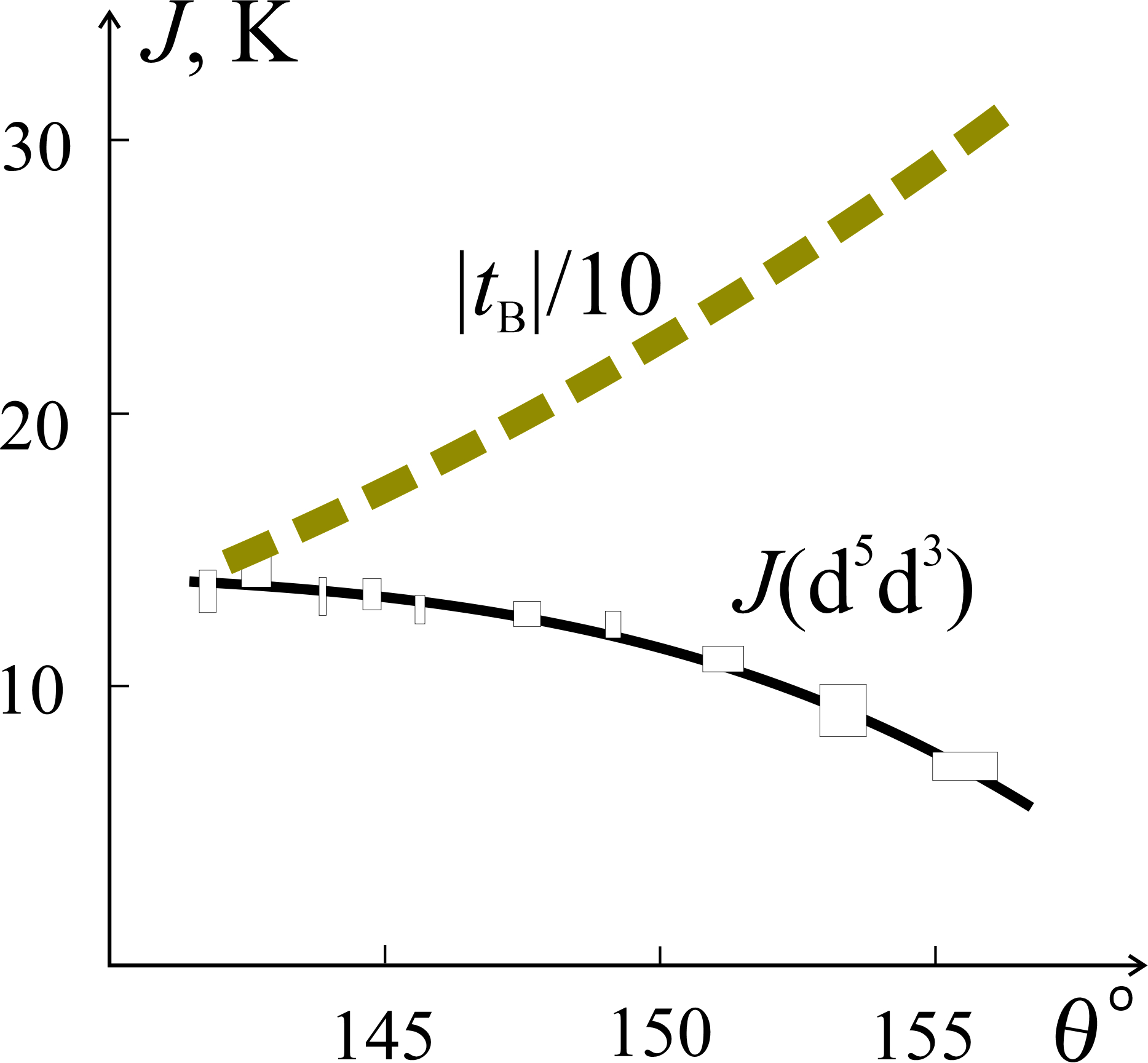}
\caption{(Color online) Angular dependence of $J(d^5d^3)$ and  $\frac{1}{10}|t_B|$ which define the effective integral $J_{eff}=J(d^5d^3)-0.1|t_B|$ }
\label{fig2}
\end{center}
\end{figure}

 Figure\,\ref{fig2} shows the dependence of the superexchange integrals $J_{FeCr}$\,=\,$J(d^5d^3)$ and  $J(e_ge_g)$  on the cation-anion-cation superexchange angle, typical for orthoferrites and orthochromites. The empty rectangles for $J(d^5d^3)$ reproduce the experimental data\,\cite{Ovanesyan} taking into account the measurement errors of the exchange integrals and the average values of the superexchange bonding angles.
 Dashed curve in Fig.\,\ref{fig2} describes the angular dependence (\ref{kinetic}) for $J(e_ge_g)$ with quantitative estimates based on the analysis of the full set of experimental data on the value of exchange parameters for orthoferrites and orthochromites\,\cite{Ovanesyan,MC_2021,JETP_2021}.

 The fitting allows us to predict the sign change for  $J_{FeCr}$ at $\theta_{cr}$\,$\approx$\,160-170$^{\circ}$. In other words, the ($t_{2g}^3e_g^2-O^{2-}-t_{2g}^3$) superexchange coupling becomes ferromagnetic at $\theta \geq \theta_{cr}$.

At variance with $J(d^5d^3)$ the exchange parameter $J(e_ge_g$)\,$\approx  |t_B|$ is shown to rapidly fall  with the decrease in the bonding angle $\theta$, so that at $\theta_{cr}$\,$\approx$\,142$^{\circ}$ we come to compensation for the ferro- and antiferromagnetic contributions to the effective exchange parameter $J_{eff}=J(d^5d^3)-0.1|t_B|$ with $S$\,=\,1,2,3,4 degeneracy and a dramatic $S$\,=\,4$\rightarrow$ $S$\,=\,1 transformation of the spin ground state with tenfold reduction in the effective transfer integral of the composite boson (see Exp.\,(\ref{tB})).

We believe that the results of the analysis of the angular dependence of the parameters $J(d^5d^3)$ and $J(e_ge_g)$, presented in Fig.\,\ref{fig2}, can be used to analyze the spin structure of EH-dimers in JT magnets with a perovskite structure such as manganites, ferrates, and ruthenates (see Table\,1).

So, for example, for the superexchange geometry, typical for
   LaMnO$_3$\,\cite{Alonso} with the Mn-O-Mn bonding angle
   $\theta \approx 155^{\circ}$ we find $J(d^5d^3)\approx$\,+7\,K and $J(e_ge_g)\approx |t_B|\approx$\,297\,K. In other words, for the effective exchange integral $J_{eff}$ we come to a rather large value: $J_{eff}=J(d^5d^3)-0.1|t_B|\approx $\,-23\,K.  Despite the antiferromagnetic sign of the Heisenberg superexchange integral these data unambiguously point to a dominant ferromagnetic contribution of the bosonic double exchange mechanism with the ground ferromagnetic $S$\,=\,4 spin state for EH-dimer and maximal "nonreduced"\, value of the composite boson transfer integral.


    For the bonding angle $\theta =143^{\circ}$ typical for the heavy rare-earth manganites RMnO$_3$ (R=Dy, Ho, Y, Er)\,\cite{Alonso} the relation between $|t_B|\approx$\,154\,K and $J(d^5d^3)\approx$\,14\,K\,\cite{Ovanesyan} approaches to the critical one: $|t_B|=10\,J(d^5d^3)$ evidencing a destabilization of the ferromagnetic state for the EH-dimers.

Thus,  the structural factor plays a significant role for stabilization of one or another spin state of the EH-dimers and effective transfer integral for the composite boson.
We believe that the change (decrease) in the angle of the cation-anion-cation superexchange bond with suppression of ferromagnetic interaction and metallicity can be the main reason for the strong effect of the substitution of Sr by Ca in JT magnets such as SrFeO$_3$, SrRuO$_4$, Sr$_2$RuO$_4$, and  Sr$_3$Ru$_2$O$_7$.

\section{Possible phase states of JT-magnets with instability to charge transfer}

In the limit of strong electron correlations and the prevalence of potential energy for valence electrons the charge-transfer-resistant stable "progenitor"\, JT-systems, as a rule, will be spin-magnetic insulators with a certain orbital ordering (OO) as a consequence of the cooperative JT effect. In the opposite limit of weak correlations and the prevalence of kinetic energy for valence electrons, we arrive at a system of itinerant electrons that form a Fermi liquid.

In  the crossover CT-instability regime, instead of a single inactive charge $d^n$-component, the on-site Hilbert space of the $d$-centers includes a charge triplet of $d^n, d^{n\pm 1}$-centers, which leads to the appearance of at least the eight  parameters of diagonal and off-diagonal charge order\,\cite{ASM_JMMM_2022}. Taking into account the spin degree of freedom and lattice modes, we arrive at
a huge variety of possible phase states. The phase diagram’s complexity originates from the specific crystal chemistry and a fine balance between the energies of the electron-lattice interaction, crystal field, local (Coulomb and exchange, or Hund) and nonlocal charge correlations, inter-site single and two-particle (composite boson) charge transfer, and spin-spin exchange. An inevitable consequence of the competition of many order parameters will be phase separation and the possibility of fine tuning of physical properties by changing the chemical composition, applying external pressure, and going over to epitaxial films and heterostructures.

Taking into account the coexistence of one- and two-particle transport, the high-temperature disordered phase for such systems will be a kind of "boson-fermion soup"\,\cite{Pepin}, or a "strange/bad"\, metal with a $T$-linear resistance dependence (strange metal) and a violation of the Mott-Ioffe-Regel criterion (bad metal). Indeed, the "strange/bad"\, metal behavior is common to practically all the JT-magnets listed in Table\,I.

One or another long-range order in JT magnets starts to form at high temperatures in a disordered phase, which is characterized by competition between the electron-lattice interaction, spin and charge fluctuations in the "struggle" for the low-temperature ground state. The local JT interaction leads to the stabilization of low-symmetry insulating magnetic structures. Low-energy charge fluctuations of the type of local anti-JT disproportionation reaction (\ref{dis}) depending on the ratio between the parameters of local and non-local correlations, the integrals of one- and two-particle transfer, and also the parameters of the electron-lattice interaction with a specific for electron-hole pairs breathing  mode can lead to the formation of a wide variety of phases from charge (CO) and spin-charge ordering, collinear and noncollinear magnetic ordering, a coherent metallic Fermi-liquid FL phase, a bosonic superconductivity (BS) phase, and also  specific nematic phase with the EH-dimer ordering\,\cite{ASM_JMMM_2022,FTT-2020}.



We believe that the expected superconductivity of JT-magnets is not a consequence of the BCS-type pairing, but the result of a  quantum transport of the effective on-site composite electron/hole bosons.
The superconducting state, as one of the possible ground states of JT magnets, can compete with the normal Fermi-liquid state, charge order, spin-charge density wave, collinear or noncollinear magnetic order, as well as specific quantum phases. The variety of competing phases clearly indicates the important role of phase separation effects\,\cite{ASM_JMMM_2022,ASM_JPCS_2022}, which must be taken into account first of all when analyzing experimental data.

Below, without dwelling on a detailed analysis of phase states and phase diagrams, we consider only the main features of the single- and two-band JT-magnets in fully disproportionated state, when they form a system of spin-singlet or spin-triplet composite bosons in a nonmagnetic or magnetic lattice, respectively.
Strictly speaking, to describe disproportionate systems, it is necessary to take into account the electron-lattice interaction, primarily with the so-called breathing mode, but below we will consider the effective Hamiltonian of effective composite bosons in the "frozen"\ lattice approximation.


\section{Single-band JT-magnets}


\subsection{Effective Hamiltonian of a system of spin-triplet composite bosons: non-magnetic lattice}

As can be seen from Table\,1 the anti-Jahn-Teller disproportionation in the system of tetrahedral JT centers with the configuration 3$d^1$, 4$d^1$, low-spin octa-centers with the configuration 3$d^7$, 4$d^7$ or octa-centers with configuration 3$d^9$, 4$d^9$ leads to the formation of a half-filled system of effective  spin-triplet bosons  moving in a non-magnetic lattice.
We represent the Hamiltonian of such a system in the form as follows
\beq
\mathcal{H} = - \sum_{i>j,\nu}t_{ij}
\left( {\hat B}^{\dag}_{i\nu} {\hat B}_{j\nu} + {\hat B}_{i\nu} {\hat B}^{\dag}_{j\nu}\right)
+  \sum_{i>j,\nu ,\nu^{\prime}} V_{ij} n_{i\nu} n_{j\nu^{\prime}} -\sum_{i,\nu}\mu_{\nu}n_{i\nu}
+ \mathcal{H}_s \, ,
\label{H}
\eeq
where $t_{ij}$ is the spin-independent boson transfer integral, $V_{ij}$ is effective boson-boson repulsion (nonlocal correlations), $\mu$ is chemical potential, $\mathcal{H}_s$ is spin Hamiltonian. The chemical potential $\mu$ is introduced to fix the boson concentration $n=\frac{1}{N}\sum_{i\nu}\langle \hat n_{i\nu}\rangle$.

The composite boson creation/annihilation operators ${\hat B}_{i\nu}^{\dagger}/{\hat B}_{i\nu}$, regardless of the spin component $\nu =0, \pm 1$, obey the on-site anticommutation Fermi relations and  the inter-site Bose commutation relations:
\beq
\{{\hat B}_i, {\hat B}^{\dag}_i\}=1\,\,, [{\hat B}_i, {\hat B}^{\dag}_j]= 1 \, .
\eeq
The anticommutation Fermi relations can be rewritten as
\beq
[{\hat B}_i, {\hat B}^{\dag}_i]=1-2{\hat B}^{\dag}_i{\hat B}_i=1-2{\hat N} _i\, .
\eeq
On the whole, these relations rule out the on-site double filling.

To take into account the influence of an external magnetic field, one can
 use the standard Peierls substitution
\begin{eqnarray}
t_{ij} \rightarrow t_{ij}e^{i(\Phi _{j}-\Phi _{i})},
\end{eqnarray}
with
\begin{eqnarray}
(\Phi _{j}-\Phi _{i})=-\frac{q}{\hbar c}\int _{{\bf R}_{i}}^{{\bf R}_{j}}{\bf A}({\bf r})d{\bf l},
\end{eqnarray}
where ${\bf A}$ is the vector potential of a homogeneous magnetic field, the integration goes along the line connecting the sites $i$ and $j$.
 In the general case spin Hamiltonian $\mathcal{H}_s $ for the system of spin-triplet bosons can be represented as follows
\beq
\mathcal{H}_s=\sum_{i>j}J_{ij}\left(\hat {\bf s}_i\cdot\hat {\bf s}_j\right) + \sum_ {i>j}j_{ij}\left(\hat {\bf s}_i\cdot\hat {\bf s}_j\right)^2 + K_{SIA}\sum_i\left({\bf m}_i\cdot \hat {\bf s}_i\right)\left({\bf n}_i\cdot\hat {\bf s}_i\right)+V_{TIA} -\sum_i\left({\bf h}\cdot\hat{\bf s}_i\right) \,,
\label{H0}
\eeq
where $J_{ij}$ and $j_{ij}$ are the bilinear and biquadratic isotropic exchange integrals, respectively, $K_{SIA}$ is a constant, ${\bf m}$ and ${\bf n}$ are unit vectors that define two characteristic axes of the second-order single-ion anisotropy,  $V_{TIA}$ is two-ion bilinear and biquadratic anisotropy, ${\bf h}$ is external field.

It is worth noting that  the Cartesian form of the composite boson spin operator can be represented as follows
\beq
{\hat s}_{\beta}={\hat B}^{\dag}_{\alpha}\epsilon_{\alpha\beta\gamma}{\hat B}_{\gamma} \, ,
\eeq
where $\epsilon_{\alpha\beta\gamma}$ is Levi-Civita tensor, $\alpha , \beta , \gamma =x, y, z$.

In the paramagnetic region, Hamiltonian (\ref{H0}) actually reduces to the Hamiltonian of the  well known lattice hard-core ($hc$) Bose system with an inter-site repulsion,  governed in the nearest-neighbor approximation only by two parameters, $t_B$ and $V$. At half-filling, depending on the relative values of the parameters, we arrive at a charge order (CO) or Bose-superfluid (BS) phase.
As the temperature decreases, one or another magnetic order is realized in the system.

\subsection{$d^1$, $d^3$ JT-magnets}

Practically the only JT-magnets known in the literature with tetrahedral $d^1$-centers, such as $\beta$-Sr$_2$VO$_4$ with V$^{4+}$ and (Sr,Ba)$_3$Cr$_2$O$_8$ with Cr$^{5+}$, are considered to be typical insulators exhibiting Jahn-Teller distortions with orbital ordering and the formation of a system of weakly coupled spin dimers (see, e.g., Refs.\,\cite{betaSr2VO4,Eremin,Cr5+}).
We did not find any literature data on JT-magnets with tetrahedral $d^3$-centers, except for the assumption made in Ref.\,\cite{melilite} about the possibility of synthesizing Ba$_2$VGe$_2$O$_7$ melilite with V$^{2+}$ ions, an anticipated JT-multiferroic.

\subsubsection{$d^7$ JT-magnets}

The origin of the metal-to-insulator transition (MIT) in the series of rare-earth nickelates RNiO$_3$ with perovskite structure has challenged the condensed matter research community for almost three decades\,\cite{RNiO3}. Furthermore, the recent theoretical prediction for superconductivity in LaNiO$_3$ thin films\,\cite{Ni-HTSC} has also triggered intensive research efforts.

The complex MIT phenomena in these materials are a perfect illustration of the competition
between the potential and kinetic energy gain, presumably governed by structural factor, namely, the Ni-O-Ni bond angle, and clear evidence for strong electron-lattice effects, which have a dramatic effect on the character of the MIT.


Orthorhombic RNiO$_3$ (R =Pr,... Lu) exhibit a first order metal-insulator phase transition  to a charge ordered  insulating state upon cooling below T$_{CO}$\,=\,T$_{MIT}$ spanning from 130\,K for Pr to $\sim$\,550-600\,K for heavy rare earths\,\cite{RNiO3}.  All these exhibit clear signatures of the charge disproportionated state with two types of Ni centers corresponding to alternating large [NiO$_6$]$^{10-}$ (Ni$^{2+}$ center) and small [NiO$_6$]$^{8-}$ (Ni$^{4+}$ center) octahedra  strongly differing in magnetic moments ($\sim$\,2\,$\mu_B$ and  $\sim$\,0, respectively) in full accordance with the disproportionation model (see Table\,I).
At low temperatures ortho-nikelates show a magnetic phase transition toward an unusual
antiferromagnetic structure defined by a  propagation vector
(1/2,0,1/2)\,\cite{RNiO3}, which can be explained by a rather strong superexchange $nnn$ (next-nearest neighbor) coupling of magnetic $S$\,=\,1 Ni$^{2+}$ centers.
The largest anomaly at T$_{MIT}$\,=\,T$_N$\,130\,K in PrNiO$_3$ was observed in the amplitude of the breathing mode, which undergoes a sharp jump of 0.15\,\AA\,\cite{Gavrilyuk}. A further interesting observation is the existence of a nearly perfect linear correlation between the amplitude of the breathing mode associated to the charge order and the staggered magnetization below the MIT. In addition, the authors\,\cite{Gavrilyuk} suggest the existence
of a hidden symmetry in the insulating phase, which may be related to a nematic contribution of bound EH dimers.

Increasing the Ni-O-Ni  bond angle when moving from LuNiO$_3$ to LaNiO$_3$ leads to a gain in kinetic energy with a clear trend to a metallization due to two important effects, namely, an increase in the transfer integrals for the $e_g$-electrons and a decrease in the parameter $V$ of inter-site repulsion (nonlocal correlations) due to an increase in the Ni-Ni separation.
 So, the x-ray diffraction, neutron scattering, transport, and thermodynamic experiments showed that globally rhombohedral single crystal LaNiO$_3$ samples  revealed a puzzlingly high metallicity, paramagnetic behavior down to 1.8\,K \,\cite{LaNiO3_1} or some signatures of antiferromagnetic transition at 157\,K\,\cite{LaNiO3_2} but no structural and metal-insulator transitions.
Combined total neutron scattering and broadband dielectric spectroscopy experiments on polycrystalline samples\,\cite{LaNiO3_3} indicated that the structure of LaNiO$_3$ has a high degree of symmetry when viewed on long length scales, but similar to orthorhombic nickelates also has at least two different types of Ni sites when viewed locally. LaNiO$_3$ is locally distorted to orthorhombic at room temperature, and further to monoclinic at 200\,K from globally rhombohedral structure\,\cite{LaNiO3_4}.   This controversial behavior for LaNiO$_3$ can be the result of the peculiar "ortho-mono-rhombo"\, phase separation.

Another example of nickel JT-magnets is the  quasi-2D nickelates ANiO$_2$ (A\,=\,Ag, Li, Na) which reveal  existence of unconventional ground states stabilized by the frustrated triangular lattice geometry  from a cooperative JT
ordering   of Ni$^{3+}$ ions in NaNiO$_2$ to a {\it moderately charge ordering} 3Ni$^{III+}$\,$\rightarrow$\,Ni$^{2+}$+2\,Ni$^{3.5+}$ in antiferromagnetic metal AgNiO$_2$\,\cite{AgNiO2}. In the case of LiNiO$_2$ there could be a competition between charge and orbital ordering, the nickel valency could be a mixture of 2+, 3+, and 4+\,\cite{LiNiO2}. Compatison of NaNiO$_2$ with LiNiO$_2$, where a number of different possible ground states are very close in energy, illustrates how two systems which are apparently so similar chemically, can nevertheless have very different behaviour\,\cite{LiNiO2}.



\subsection{$d^9$ JT-magnets}

\subsubsection{Isoelectronic quasi-2D cuprates and nickelates}


The Cu$^{2+}$ ion in octahedral complexes is characterized by the strongest JT bond and is the most popular, almost "textbook"\, illustration of the Jahn-Teller effect. The consequence of this effect is the formation of the dielectric state of a quantum antiferromagnet, for example, in KCuF$_3$ and La$_2$CuO$_4$ or quasi-2D ferromagnet K$_2$CuF$_4$. However, in contrast to fluorides, in La$_2$CuO$_4$ the JT distortion leads to the formation of CuO$_2$-planes with a "perovskite"\ configuration of CuO$_4$-clusters with the ground $b_{1g}\propto d_{x^2-y^2}$ state of the $e_g$-hole, which provides a strong $\sigma$-coupling channel for hole transfer in the CuO$_2$ plane and disproportionation (\ref{disCuO}) to form spin- singlet and orbitally nondegenerate (${}^1A_{1g}$) electronic [CuO$_4$]$^{7-}$ (analog  of Cu$^+$ ion) and Zhang-Rice (ZR)\,\cite{ZR} hole [CuO$_4$]$^{5-}$ (analog of the Cu$^{3+}$ ion) centers.


Recently\,\cite{ASM_JMMM_2022,OS_2023} we argued that there are no fundamental qualitative differences in the electronic structure of "apex-free"\, RNiO$_2$ nickelates and cuprates, primarily cuprates with a $T^{\prime}$-structure. The unusual properties of cuprates and nickelates are the result of "competition"\ of various parameters that govern the ground state of the CuO$_2$ (NiO$_2$) planes. Thus, if for the vast majority of parent cuprates an antiferromagnetic dielectric phase is observed, which corresponds to the limit of strong local correlations, then this phase was not found in the parent nickelates RNiO$_2$, which can be associated with a smaller value or even a change in the sign of the local correlation parameter. We have proposed\,\cite{ASM_JMMM_2022,OS_2023} to understand by "parent"\ cuprate or nickelate with hole half-filling of in-plane centers CuO$_4$ (NiO$_4$), which, depending on the parameters of local and non-local correlations, transfer integrals , exchange integrals, as well as the "external"\, crystal field formed by the out-of-plane environment, can have a different ground state - an antiferromagnetic insulator (AFMI), an unusual Bose superconductor (BS), a Fermi metal (FL) or a non-magnetic insulator with a charge ordering (CO). Obviously, these phases will differ not only in electronic but also in lattice degrees of freedom, the interaction of which ensures the minimum of the total free energy. In addition, the competition of several possible phases with similar energies will lead to phase separation, which will have a significant effect on the observed physical properties.


To describe the actual low-energy phase states of cuprate-nickelates, we proposed a minimal model for the CuO$_2$/NiO$_2$-planes with the on-site Hilbert space reduced to a charge triplet of the three effective valence centers [CuO$_4$]$^{5-,6-,7-}$/[NiO$_4$]$^{6-,7-,8-}$ (nominally Cu$^{3+,2+,1+}$/Ni$^{2+,1+,0+}$) with different conventional spin, different orbital symmetry, and different local lattice configuration\,\cite{truegap,Moskvin-JSNM-2019,EH,FMM,ASM_CM_2021,ASM_JMMM_2022,OS_2023}.
Making use of  the $S$\,=\,1 pseudospin formalism and the spin-pseudospin operators of the type of the Hubbard $X$-operators we have constructed the spin-pseudospin Hamiltonian of the charge triplet model, which takes into account local and nonlocal correlations, correlated one-particle and two-particle (bosonic) transport, and Heisenberg spin exchange. In particular cases the Hamiltonian  reduces to the well-known "limiting" Hamiltonians (Hubbard, Heisenberg, atomic limit, hard-core bosons, ...). In accordance with experimental data for apexless cuprates\,\cite{Naito}, nickelates\,\cite{Li}, and different typical cuprates we argue that antiferromagnetic insulating (AFMI), charge ordered (CO), Bose superconducting (BS), and Fermi-liquid (FL) phases are possible phase states of a model parent cuprate/nickelate, while typical phase states of  doped systems, in particular, mysterious pseudogap phase, are the result of a phase separation (PS). Superconductivity of cuprates/nickelates is not a consequence of pairing of doped holes\,\cite{Pelc}, but the result of quantum transport of on-site composite hole bosons, whereas main peculiarities of normal state can be related to an electron-hole interplay for unusual Fermi-liquid phase and features of the PS. Puzzlingly, but it is the electron-lattice interaction, which in the BCS model determines $s$-wave pairing, in the model of local composite bosons gives $d_{x^2-y^2}$-symmetry of the superconducting order parameter, thus showing once again a substantial involvement of the lattice in the HTSC. Within the framework of the effective field approximation, and the Maxwell construction, we have constructed a number of 2D  T–$p$ phase diagrams for the CuO$_2$/NiO$_2$ planes, which qualitatively reproduce main features of the experimentally observed 3D phase diagrams of cuprates and nickelates\,\cite{ASM_JMMM_2022} (see Fig.\,3). The pseudogap phase is associated with the PS region AFMI-CO-FL-BS, separated from the 100\% FL-phase by the curve T$^*(p)$ of the "third order"\, phase transition.

 \begin{figure*}[t]
\centering
\includegraphics[width=13.5 cm,angle=0]{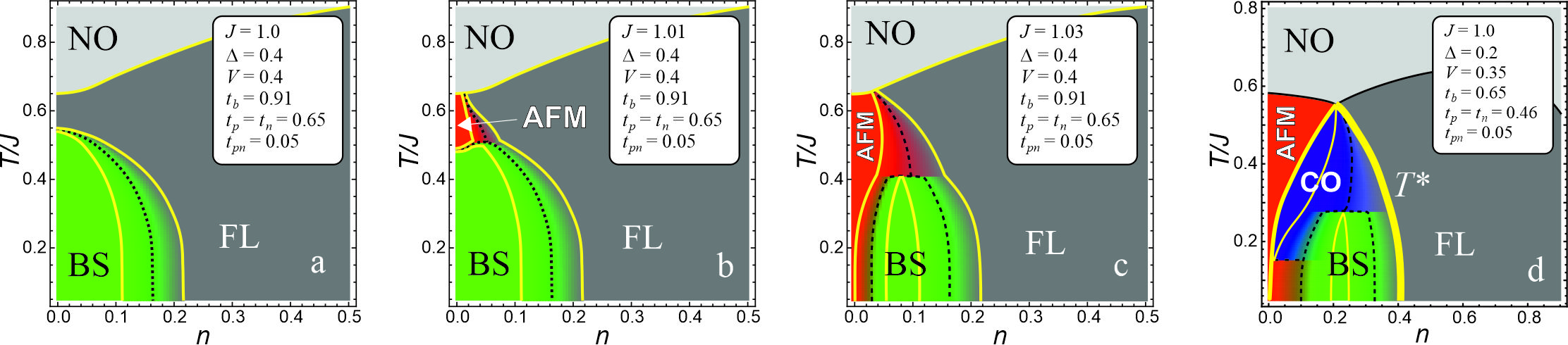}
\caption{(Color online) Model phase $T$\,-\,$n$-diagrams of hole-doped CuO$_2$/NiO$_2$ planes in cuprates/nickelates calculated in the effective field approximation ($n=p$ for hole doping) with the phase separation taken into account using Maxwell's construction; $J$ is exchange integral, $\Delta =U/2$ local correlation parameter, $V$  nonlocal correlation parameter, $t_p, t_n, t_{pn}$ are three independent integrals of  correlated single-particle transfer, $t_B$ is the effective transfer integral of the composite boson (see insets) assuming competition between "monophases"\, NO (disordered), AFMI, BS, FL, CO. The boundaries between the phases represent lines of equal free energies.  The dashed curves on ($a$)--($d$) indicate the line of equal volume fractions of two neighboring phases, the yellow curves  represent the lines of phase transitions of the "third"\ kind, limiting the regions with maximal 100\% volume fraction of one of the phases.  See Ref.\,\cite{ASM_JMMM_2022} for detail. Let us pay attention to the strong change in the phase diagram even with a very small change in the parameters of the Hamiltonian (compare panels $a$, $b$, $c$).}
\label{fig3}
\end{figure*}

In general, quasi-2B cuprates and nickelates provide an excellent example of the applicability of the anti-JT disproportionation model.
A large amount of experimental data from a long-term study of various properties of a wide class of old 2D cuprates and novel 2D nickelates, as well as the results of theoretical modeling of phase diagrams in the charge triplet model\,\cite{ASM_JMMM_2022} provide important information about possible phase states of other JT magnets  with  charge transfer.

\subsubsection{"Silver"\, JT-magnets}

The anti-JT disproportionation model predicts the possibility of a "silver or palladium path"\, to superconductivity in systems based on Ag$^{2+}$(4d$^9$) or Pd$^{+}$(4d$^9$), that is, the 4d analog of Cu$^{2+}$. The most likely candidate, silver fluoride AgF$_2$\,\cite{AgF2,AgF2_1,AgF2_2}, also known as $\alpha$-AgF$_2$, is an excellent analogue of cuprate with surprisingly close electronic parameters to La$_2$CuO $_4$, but with greater deformation (buckling) of AgF$_2$ planes. However, this fluoride is a canted antiferromagnetic insulator, although close to a charge-transfer instability. Indeed, experimental studies\,\cite{AgF2_beta} report the discovery of a metastable disproportionated diamagnetic phase $\beta$-AgF$_2$, interpreted as a charge-ordered compound Ag$^{1+}$Ag$^{3+}$F $_4$, which quickly transforms into the $\alpha$-AgF$_2$ structure.

Unlike the antiferromagnetic insulator Cu$^{2+}$O, its silver 4d analogue Ag$^{2+}$O is a diamagnetic semiconductor with a disproportionate Ag sublattice, whose chemical formula is often written as Ag$^{1 +}$Ag$^{3+}$O$_2$ with O-Ag$^{1+}$(4d$^{10}$)-O collinear bonds and Ag$^{3+}$ square planar bonds (4d$^{8}$)O$_4$\,\cite{AgO,AgO_1}. In this case, the [AgO$_4$]$^{5-}$ cluster, like the [CuO$_4$]$^{5-}$ center in cuprates, is in a nonmagnetic state of the Zhang-Rice singlet type.

\section{Two-band JT magnets }


Single-band JT magnets, with their relatively simple electronic structure, provide an excellent illustration of the predictive power of the anti-JT disproportionation model, while the situation with two-band JT magnets is less certain.

Anti-Jahn-Teller disproportionation in "two-band"\, systems of high-spin octa-centers with 3$d^4$, 4$d^4$ configuration or tetrahedral JT centers with 3$d^6$, 4$d^6$ configuration implies unusual phase with the coexistence of a half-filled system of effective spin-triplet electron or hole bosons with  configuration $e_g^2:{}^3A_{2g}$ or $\underline{e}_g^2:{}^3A_{2g}$ and  a magnetic lattice with the on-site $S=3/2$ configurations $t_{2g}^3:{}^4A_{2g}$, although this does not exclude the existence of unusual phases with delocalized $t_{2g }$-electrons (see  review article\,\cite{Hirschfeld}).

Two-band JT magnets include a large number of promising compounds, most of which are presented in the last column of Table\,I. Below, we briefly consider effective Hamiltonian, the features of the electronic structure and physical properties of the most prominent representatives of two-band JT magnets.


\subsection{Effective Hamiltonian of a system of spin-triplet composite bosons: magnetic lattice  }


Anti-Jahn-Teller disproportionation in a system of high-spin octahedral  JT-centers with 3d$^4$, 4d$^4$ configurations or tetrahedral JT-centers with 3$d^6$, 4$d^6$ configurations leads to the formation of a half-filled system of effective spin-triplet electron or hole bosons with  configuration $e_g^2:{}^3A_{2g}$ or $\underline{e}_g^2:{}^3A_{2g}$, moving in a magnetic lattice with the on-site configurations $t_{2g}^3:{}^4A_{2g}$ (see Table\,I).

The effective Hamiltonian of such a system can also be represented as (\ref{H}), however, with the spin-dependent composite boson transfer integral (see Exp.\,(\ref{tB}))
\begin{equation}
	t_{ij}=\frac{S(S+1)}{20}t_B\,,
 \label{tBS1}
\end{equation}
where $\hat{\bf S}=\hat{\bf S}_i+\hat{\bf S}_j$ is the total spin of the EH-pair ($ij$), $S$\,=\,1, 2, 3, 4.

In contrast with the single-band  JT-magnets the spin Hamiltonian $\mathcal{H}_s $ for two-band JT-magnets will have a much more complex structure. Taking into account only the bilinear spin-spin isotropic exchange, it can be represented as follows
\begin{equation}
\mathcal{H}_s =
\sum_{i>j}J_{ij}^{ll}(\hat{\bf S}_{i}\cdot\hat{\bf S}_{j})+
+\sum_{i>j}J_{ij}^{bb}(\hat{\bf s}_{i}\cdot\hat{\bf s}_{j})
+\sum_{i>j}J_{ij}^{bl}(\hat{\bf s}_{i}\cdot\hat{\bf S}_{j})
+\sum_{i}J_{ii}^{bl}(\hat{\bf s}_{i}\cdot\hat{\bf S}_{i})\,,
\label{HH}
\end{equation}
where we assume the localized $t_{2g}$ subshell. The first term describes the exchange interaction of the "lattice" spins, the second term describes the exchange interaction between spin-triplet bosons, the third and fourth terms describe the exchange between bosons and lattice spins, and the last term actually describes the intra-atomic Hund exchange.
To fulfill Hund's rule, it is necessary to set the exchange integral $J^{bl}_{ii}$ to be relatively large ferromagnetic.




Estimates for different superexchange couplings given the cation-anion-cation bond geometry typical for perovskites such as ferrates (Ca,Sr)FeO$_3$ or manganites RMnO$_3$ with bare octa-HS d$^4$ configurations\,\cite{Moskvin-09} predict antiferromagnetic coupling for the $nn$ lattice centers ($J^{ll}>0$)  and for the two nearest neighbor bosons  ($J^{bb}>0$), whereas the coupling of the boson and the nearest neighbor lattice centers ($J^{bl}<0$) can be ferro- or antiferromagnetic depending on the value of the cation-anion-cation bonding angle (see Fig.\,\ref{fig1}). Taking into account the boson transport which prefers an overall ferromagnetic ordering  we arrive at a highly frustrated system with a competition of the ferro- and antiferromagnetic interactions.

Generally speaking, our model Hamiltonian describes the system that can be
considered as a Bose-analogue of the {\it one orbital} double-exchange model system\,\cite{Dagotto}.


\subsection{Chromium Cr$^{2+}$ compounds}
Among the JT chromium compounds, we have more or less reliable information about chromium difluoride CrF$_2$, according to which it is an antiferromagnetic insulator\,\cite{CrF2_AFM}. However, X-ray absorption and resonant inelastic x-ray scattering (RIXS) spectra of CrF$_2$\,\cite{CrF2} point to a presence of three chromium oxidation states, namely Cr$^+$, Cr$^{2+}$, and Cr$^{3+}$, which indicates  instability with respect to charge transfer with   clear signatures of the $d$-$d$ disproportionation reaction in this JT magnet, The most likely explanation for this puzzle is phase separation, that is, the coexistence of antiferromagnetic regions and regions of a disproportionate phase.

\subsection{Manganites RMnO$_3$}

Features of the anti-JT disproportionation and its influence on the phase diagram of manganites RMnO$_3$ are considered in detail in Ref.\,\cite{Moskvin-09}.

A high-temperature thermally fluctuating charge disproportionated metallic state has been postulated for LaMnO$_3$ by different authors\,\cite{LaMnO3_1,LaMnO3_2,Good}. However, actually upon lowering the temperature one observes a first order phase transition at T=T$_{JT}$ (T$_{JT}$\,$\approx$\,750\,K in LaMnO$_3$) from the high-temperature  fully disproportionated Bose metallic  phase to a low-temperature orbitally ordered  insulating   phase with a cooperative Jahn-Teller ordering of the occupied $e_g$-orbitals of the Mn$^{3+}$O$_6$ octahedra accompanied by A-type antiferromagnetic
ordering below T$_N$ (T$_{N}$\,$\approx$\,140\,K in LaMnO$_3$)\,\cite{Good,Moskvin-09}.  However, many experimental data point to a phase separation with the coexistence of insulating and disproportionated phases\,\cite{Moskvin-09,Ritter}.

The nonisovalent substitution and/or nonstoichiometry seems to revive the disproportionated  phase and such manganites along with a metallic ferromagnetism with a colossal magnetoresistance reveal many properties typical for local spin-triplet superconductivity\cite{Moskvin-09,Kim,Krivoruchko,Markovich,Kasai,Mitin,Muk}.

Distinct signatures of high-temperature disproportionated  phases are revealed also in other manganites, such as  LaMn$_7$O$_{12}$\,\cite{LaMn7O12} with quadruple perovskite
structure and YBaMn$_2$O$_6$\,\cite{YBaMn2O6}.



\subsection{Iron Fe$^{4+}$ JT magnets}

All the ferrates listed in Table\, I, one way or another, are JT magnets that are unstable with respect to charge transfer.

 The AFe$^{4+}$O$_3$ (A = Ca, Sr, Ba) perovskites show intriguing physical properties which are strongly dependent on the size and polarizability  of the A-site ion since this affects all the main parameters governing their electronic structure.

With decreasing temperature orthorhombic metallic CaFeO$_3$  (CFO) exhibits a second-order phase transition to a narrow-gap semiconducting charge-ordered monoclinic semiconductor, or Hund’s insulator,  with disproportionation into Fe$^{4\pm\delta}$  below a transition temperature  T$_{CO}$\,=\,T$_{MIT}$\,=\,290\,K at ambient pressure, resulting in a three-dimensional rock salt type ordering of alternating small and large oxygen octahedra surrounding the nominal $d^3$ and $d^5$ Fe sites, respectively\,\cite{CFO}. Parameter $\delta$=0 for T\,>\,290\,K, it increases continuously with decreasing temperature below 290\,K; typically $\delta$ approaches unity at low temperatures. The MIT is accompanied by the reduction of crystal symmetry as well as the sharp variation in electrical transport.
Within our model the disproportionated phase in CFO implies the electron boson confinement in the larger FeO$_6$ octahedra.

The charge-disproportionation scenario for CFO  has been well established experimentally using ${}^{57}$Fe M{\"o}ssbauer spectroscopy\,\cite{Mossbauer,CFO_1}, which clearly  reveals two different sites with considerably different isomer shifts and hyperfine fields.

Let us pay attention to the possibility of the formation of domains in the charge-ordered  state with 180$^{\circ}$-domain walls, realizing the transition between two types of "site-centered"\, charge order. At the center of the domain walls, a system of delocalized spin-triplet composite bosons with a "bond-centered" charge order is formed, which formally corresponds to the system of Fe$^{4+}$ centers.

As temperature is lowered further, there is
another transition in CFO from the paramagnetic  to  antiferromagnetic insulator at the
N\'{e}el temperature T$_N$ $\approx$\,120\,K.   The low-temperature magnetic data can be fit equally well by a screw spiral structure or by a sinusoidal amplitude-modulated
structure. The values of the moments at the two Fe sites are allowed to
take different values; 2.5 and 3.5 $\mu_{B}$ for the spiral structure,
and maximum amplitudes of 3.5 and 5.0 $\mu_{B}$  for the sinusoidal structure\,\cite{CFO}.

Note that the high-temperature orthorhombic metal phase of CFO can be viewed as a  Hund’s bad metal which appears as a mixed-valence state that fluctuates between two atomic configurations.

Different from the distorted perovskite CaFeO$_3$ undistorted cubic perovskites  SrFeO$_3$ and BaFeO$_3$ keep metallic behaviors down to very low temperatures with different type of helical spin order. However, the ground state in these ferrates raises a lot of questions.
At variance with M\"{o}ssbauer data for CaFeO$_3$
the single magnetic hyperfine pattern for SrFeO$_3$ at 4\,K indicates a rapid electron exchange between Fe$^{3+}$ and Fe$^{5+}$ ions, for the center shift and the hyperfine field coincide approximately with the average values of the corresponding parameters for CaFeO$_3$\,\cite{Mossbauer}. In other words, "static"\, disproportionation occurs in CaFeO$_3$ with the formation of a site-centered charge order, whereas in SrFeO$_3$ we are dealing with "dynamic"\, disproportionation with the formation of a bond-centered charge order.
Furthermore, experiments  have revealed in SrFeO$_3$ a phase-separated state with a surprising variety of magnetic incommensurate helical and commensurate structures\,\cite{SFO_PS}.

Surprisingly, a ferromagnetic ground state is found in BaFeO$_3$ single crystalline thin films with saturation magnetization and Curie temperature of 3.2\,$\mu_{B}$/formula unit and 115\,K, respectively\,\cite{BFO}. Unusually, for a uniform cubic ferromagnet, the films are insulating with an optical gap of $\sim$\,1.8\,eV.

An incommensurate helicoidal spin ordering observed both in CaFeO$_3$ and SrFeO$_3$\,\cite{CaSrFeO3} up to very low temperatures  can be explained  as a result of a competition between conventional exchange coupling and the bosonic double exchange. Obviously, the theoretical and experimental study of the phase diagram for (Ca,Sr)FeO$_3$ and substituted systems deserves further work, especially, aimed to a search of a possible superconductivity.

The $^{57}$Fe M\"{o}ssbauer measurements for the double-layered perovskite ferrate Sr$_3$Fe$_2$O$_7$ indicate the charge disproportionation and the magnetic properties, which are similar to CaFeO$_3$\,\cite{SFO_327}. The critical temperature for the charge disproportionation reaction and the N\'{e}el temperature T$_N$ of the helical spin order are determined to be $\sim$\,343\,K and $\sim$\,120\,K, respectively. Above 343\,K, spectra clearly show a Fe$^{4+}$ singlet. Puzzlingly, the spatial ordering pattern of the disproportionated charges has remained “hidden”\, to conventional diffraction probes, despite numerous x-ray and neutron scattering studies. Only relatively recently making use of neutron Larmor diffraction and Fe K-edge resonant x-ray scattering Kim {\it et al.}\,\cite{327_hidden order} demonstrated checkerboard charge order in the FeO$_2$ layers and show that the “invisibility”\, of charge ordering in Sr$_3$Fe$_2$O$_7$ originates from frustration of the interactions between neighboring layers.

Less studied quasi-2D ferrate Sr$_2$Fe0$_4$ with the K$_2$NiF$_4$ structure is a compound isotypic with the parent cuprate La$_2$CuO$_4$. It is antiferromagnetic semiconductor at ambient pressure with a N\'{e}el temperature T$_N$ of about 56\,K\,\cite{Sr2FeO4,Sr2FeO4_1}.
Over the past 30 years, the concept of the electronic structure of Sr$_2$Fe0$_4$ has changed from a Mott-type antiferromagnetic insulator similar to La$_2$CuO$_4$\,\cite{Sr2FeO4} to a insulator with negative charge-transfer energy (negative-$\Delta_{pd}$)\,\cite{Sr2FeO4_1},
The insulating ground state of Sr$_2$Fe0$_4$ is assumed to be stabilized by a hidden structural distortion similar to the charge order in the related Sr$_3$Fe$_2$O$_7$ and differs from the charge disproportionation in other Fe$^{4+}$ oxoferrates.

However, we believe that, in fact, the ground spin-charge state in this ferrate, as well as in other JT ferrates, is determined by $d$-$d$ anti-JT disproportionation. This is evidenced by the absence of a noticeable JT distortion of the FeO$_6$ octahedra, the manifestation of a phonon mode atypical for the K$_2$NiF$_4$ structure, which can be naturally associated with a breathing mode typical for $d$-$d$ disproportionation, an elliptical cycloidal spin spiral structure typical of all JT ferrates, an insulator-metal transition under high pressure\,\cite{Sr2FeO4_1},
For elucidating the details
of the ground state we need further  studies, in particular on single crystals of Sr$_2$FeO$_4$.

\subsection{JT ruthenates}

Just like Fe$^{4+}$(3$d^4$) JT ferrates, the Ru$^{4+}$(4$d^4$)-based ruthenates belong to the same family of  the Ruddlesden-Popper (A$_{n+1}$B$_n$O$_{3n+1}$) compounds. They host a rich  physics, including unconventional
superconductivity in Sr$_2$RuO$_4$, a metamagnetic ground state in Sr$_3$Ru$_2$O$_7$, insulating
antiferromagnetism in Ca$_2$RuO$_4$ and Ca$_3$Ru$_2$O$_7$, and both paramagnetic and ferromagnetic
metallic states in CaRuO$_3$ and SrRuO$_3$, respectively. Ruthenates undergo a
variety of electronic, magnetic, and orbital ordering transitions, which are tunable with chemical
doping, pressure, temperature, magnetic field, and epitaxial strain. However, their properties differ in many points from their 3d analogues.  Not least, this is due to the fact that the 4d shell of the Ru$^{4+}$ ion is more extended than the 3d shell of the Fe$^{4+}$ electronic analog, which most likely leads to an increase in the crystal splitting parameter 10$D$q, a decrease in the local correlation parameter, and an increase in the transfer integrals. As a result the Ru$^{4+}$ (4d$^{4+}$) ions tend to adopt a low-spin state or $S$\,=\,1 state because relatively large crystal fields often overpower the Hund’s rule coupling\,\cite{Ru_LS}.

In other words, in ruthenates we seemingly encounter a fine high spin -- low spin (HS -- LS) balance up to the possibility of coexistence of HS- and LS-states\,\cite{Ru_LS,Ru_LS1}. It means that by varying substitution, tuning the physical and chemical pressures, reducing the film thickness,  one can observe different quantum states  from typical for JT magnets as JT ferrates   to states typical for low-spin $t_{2g}^4$-systems with trend to a phase separation.

Practically all the layered ruthenates at low temperatures are characterized by a  robust Fermi liquid behavior  evidenced both by the quadratic temperature dependence of resistivity and  by the observations of quantum oscillations. However, violation of the Mott-Ioffe-Regel limit for the basal plane resistivity and anomalous strange metallic behavior with a linear temperature dependence of resistivity at high temperature clearly reveal behavior inconsistent with any conventional Fermi-liquid paradigms\,\cite{Ru_badmetal} but typical for disproportionate systems with two types of the charge transport.

Ruthenates  are an excellent candidate to explore the intricate interplay between structural and electron-spin degrees of freedom. For instance, Ca$_2$RuO$_4$ is a paramagnetic Mott insulator below the metal-insulator transition temperature T$_{MIT}\approx$\,360\,K with antiferromagnetic ordering below T$_N\approx$\,110\,K\,\cite{CaSr2RuO4}.  However, application of very modest pressures transforms it from an antiferromagnetic Mott insulator to a quasi 2D ferromagnetic metal.  Under current flow, the insulating ground state was observed to transform into an electrically conducting phase with a high diamagnetic susceptibility.

Puzzlingly, the single crystalline Ca$_2$RuO$_4$ nanofilms  exhibit co-appearance of high-temperature superconductivity with T$_c\approx$\,60\,K and ferromagnetism\,\cite{Ca2RuO4_SC}. Such a high temperature of the superconducting transition testifies in favor of an unconventional mechanism of superconductivity of the type realized in high-T$_c$ cuprates.

The replacement of Ca$^{2+}$ ions (ionic radius 1.34\,\AA) with  Sr$^{2+}$ ions (ionic radius 1.44\,\AA)  in bulk-family seemingly triggers a subtle modification in the electronic structure but a dramatic transformation of the ground state from antiferromagnetic insulating in Ca$_2$RuO$_4$ to superconducting and ferromagnetic in Sr$_2$RuO$_4$ with spiral spin structure  in the ground normal metallic state\,\cite{Maeno}.

Based on early Knight shift, polarized neutron scattering, muon-spin-resonance, and polar
Kerr measurements, Sr$_2$RuO$_4$ has been widely thought to support a spin-triplet chiral $p$-wave superconducting state\,\cite{Mack-2003}. However, despite significant achievements in characterizing the properties of Sr$_2$RuO$_4$ over the last three decades, the precise nature of its electronic ground state and superconducting order parameter is still unresolved\,\cite{Mack-2017,Legett}.
 Understanding the nature of superconductivity in Sr$_2$RuO$_4$ continues to be one of the most enigmatic problems in unconventional superconductivity  despite vast interest, as well as a wide
array of experiments performed on the material.
Recent results have pushed the community towards potentially adopting an even-parity spin-singlet pairing state, although conventional states of this nature are not able to consistently explain all observations.
  It should be noted that superconductivity turns out to be a relatively common property of ruthenates, so very recently  strain-stabilized superconductivity with T$_c\approx$\,2\,K was discovered in ruthenate RuO$_2$ films\,\cite{RuO2,RuO2_1}.

  Generally speaking, despite extensive efforts, a comprehensive understanding of electronic structure and physical properties in JT ruthenates is still lacking.

\subsection{Iron-based superconductors}

The Fe$^{2+}$ iron based superconductors have a layered structure with conducting layers made of iron which are tetrahedrally coordinated by a pnictide or chalcogenide (As, Se, S or P), which are responsible for the superconductivity. These JT magnets exhibit the unprecedented richness of the physics,  sometimes all within a single family -- magnetism, unconventional superconductivity, quantum criticality, linear-in-T resistivity, nematic order, and a tendency toward orbital selective Mottness\,\cite{Fe,Fe_2015,Fe_2016,Fe_2020}.  Researchers have found practically all phenomena associated with strongly correlated electron systems in the Fe-based materials.
 At the present time, a variety of theoretical approaches are being taken to understand these systems, although the issue remains to be fully settled.

We do not aim to give here a comprehensive review of the electronic structure and phase diagrams of iron based superconductors, but only pay attention to a number of specific features that allow us to assume an important role for the disproportionation mechanism.
Superconductivity in FePn/Ch emerges out of a “bad-metal” normal state; and the superconducting phase occurs near anantiferromagnetic order in proximity to a Mott transition. The parent iron pnictides are antiferromagnetically ordered metals, insulating behavior  and AF order also appears in a variety of iron chalcogenides.

Unconventional non-BCS superconductivity in FePn/Ch has much in common with that of the copper oxides, in particular, the ratio of T$_c$ versus the superfluid density is close to the Uemura plot as for the hole doped high-T$_c$ cuprates\,\cite{Uemura,Uemura_1}, as for cuprates the electronic nematicity has been observed in the normal state of many if not all the
FePn/Ch.

At the same time, FePn/Ch are different in many respects from the cuprates. Thus, the high field inelastic neutron scattering data in the optimally doped Fe(Se,Te) superconductor\,\cite{FeSeTe_neutron} and in 112-type pnictide\,\cite{S=1_neutron} show that similar to cuprates the magnetic fluctuations play a central role in iron superconductivity, however, these suggest that the superconductivity FePn/Ch is actually driven by a spin-triplet bound state.
The spin-triplet nature of superconducting carriers in FePn/FeCh was proposed back in 2008\,\cite{FPS,S1} and confirmed by a number of experimental facts\,\cite{S=1,S=1_STM,S=1_LiFeAs}, although experimental data are contradictory\,\cite{S_not1,Snot1}. In this regard, let us turn our attention to one of the main modern technique for determining the spin of superconducting carriers - measuring the spin susceptibility by measuring the Knight shift\,\cite{Hirschfeld}.
It is believed that spins in a triplet superconductor should be polarized in an external magnetic field, just like free spins in an ordinary metal. Thus, in such a system one can expect that the spin susceptibility and the Knight shift should not have singularities in T$_c$. Spin anisotropy can suppress this for some directions but not for others.
In a spin-singlet superconductor, the magnetic susceptibility vanishes at T$\rightarrow$\,0. Thus, for spin-singlet superconductivity, a decrease in the uniform spin susceptibility below Tc can be expected, although qualitatively the same can occur for certain components of the triplet, although the vanishing susceptibility is often difficult to determine due to the background Van Vleck contribution. However, this technique does not take into account the complex nature of spin interactions and the spin structure of  spin-triplet superconductors.

One way or another, the "singlet-triplet" dilemma for superconducting carriers in the vast majority of superconductors is considered within the framework of the BCS scenario, while the model of anti-JT disproportionation in JT magnets represents a fundamentally different view of the mechanism of superconductivity, in which superconducting carriers are effective local, singlet or triplet, hole or electronic, composite bosons.
By the way, our model assumes that superconducting carriers in FePn/Ch compounds consist of $e_g$-holes, and not of $t_{2g}$-electrons, as predicted by the single-electron multi-orbital band model\,\cite{Fe}.


At the moment, we cannot give an unambiguous conclusion about the role of the mechanism of anti-JT disproportionation in iron-based superconductors, however, in any case, finding the high-T$_c$ superconductivity in the FePn/Ch compounds with the tetrahedral coordination of the iron Fe$^{2+}$(3$d^6$) ions in HS state and its coexistence with an unconventional magnetism can be a key argument supporting the disproportionation scenario.

What seems even more surprising, our simple model gives convincing predictions of superconductivity and its features in different  quasi-two-dimensional JT magnets, cuprates, nickelates, ruthenates, and ferropnictides/chalcogenides,  differing both in the electronic structure of active centers, and in the local crystal structure. The model predicts hole-type bosonic spin-singlet superconductivity in 2D cuprates and nickelates,  spin-triplet hole superconductivity in FePn/FeCh with sufficiently high T$_c$ in both systems, and electronic superconductivity in Sr$_2$RuO$_4$ with very low T$_c$, by the way, in agreement with Hirsch's ideas about the hole nature of HTSC\,\cite{ Hirsch_holes,Hirsch}.

\section{Conclusion}
We believe that unusual properties of a wide class of JT magnets, that is materials based on Jahn-Teller 3d and 4d ions with different crystal and electronic structures, from quasi-two-dimensional unconventional superconductors (cuprates, nickelates, ferropnictides/chalcogenides, ruthenate SrRuO$_4$), manganites with local superconductivity to 3D ferrates (CaSr)FeO$_3$, nickelates RNiO$_3$ and silver oxide AgO with unusual charge and magnetic order can be explained within the framework of a single scenario, which assumes their instability with respect to anti-Jahn-Teller $d$-$d$ disproportionation.
As a result of disproportionation, the parent ("progenitor") JT magnet is transformed into a half-filled system equivalent to a single- or two-band system  of effective local composite spin-singlet or spin-triplet, electron or hole $S$-type bosons in a non-magnetic or magnetic lattice, which gives rise to an extremely rich set of phase states from non-magnetic and magnetic insulators, unusual magnetic metallic and superconducting states, to a specific nematic ordering of the EH-dimers. Effective composite bosons cannot be considered as conventional quasiparticles, they are an indivisible part of many-electron configurations. Effective spin-dependent two-particle bosonic transport  in two-band JT magnets results in a behavior typical for "double-exchange"\, systems.

The model provides a comprehensive understanding of the well established charge and magnetic order in JT  ferrates and nickelates RNiO$_3$ including nontrivial effect of the cation-anion-cation bonding angle.

The most optimal conditions for HTSC with spin-singlet local composite bosons and a spinless lattice can only be achieved for low-symmetry quasi-two-dimensional d$^9$ JT magnets, such as 2D cuprates and nickelates, where disproportionation follows the traditional Jahn-Teller effect and orbital ordering.

 The anti-JT disproportionation model points to a possibility of spin-triplet superconductivity in ruthenates Sr$_2$RuO$_4$ and RuO$_2$, ferropnictides/chalcogenides FePn/FeCh, manganite LaMnO$_3$, although in most of the known “candidates” (Ca(Sr )FeO$_3$, RNiO$_3$, AgO) one or another spin-charge order is realized. The model assumes that the effective superconducting carriers in the FePn/FeCh compounds consist of $e_g$ holes rather than $t_{2g}$ electrons, as predicted by the one-electron multi-orbital band models.  The effective Hamiltonians for spin-triplet composite bosons in nonmagnetic and magnetic lattices have a complex spin structure, which must be taken into account when interpreting experiments to determine the spin of superconducting carriers.

\acknowledgments{I thank Dr Yuri Panov for the very fruitful multi-year collaboration,  stimulating and encouraging discussions.

The research was supported by the Ministry of Education and Science of the Russian Federation, project No. FEUZ-2023-0017.}


\end{document}